\documentclass[aps,nofootinbib,showpacs,preprint]{revtex4}
\usepackage{amsmath}
\usepackage{amsfonts}
\usepackage{amssymb}
\usepackage{color}

\newcommand{\omits}[1]{}

\def\bc{\begin{center}}

\def\ec{\end{center}}
\def\be{\begin{eqnarray}}
\def\ee{\end{eqnarray}}

\definecolor{dyellow}{rgb}{1.,0.8,.0}
\definecolor{myblue}{rgb}{.1,.1,.7}
\definecolor{dcyan}{rgb}{.0,.6,.6}
\definecolor{dmagenta}{rgb}{0.6,0.0,0.6}
\definecolor{brown}{rgb}{0.6,0.2,0.}
\definecolor{darkblue}{rgb}{.0,.0,0.5}
\definecolor{darkred}{rgb}{0.75,0.0,0.0}
\definecolor{orange}{rgb}{1.,.6,.0}
\definecolor{dorange}{rgb}{0.8,.4,.0}
\definecolor{darkgreen}{rgb}{0.0,0.6,0.0}
\definecolor{purple}{rgb}{.4,.0,.4}
\definecolor{lightgrey}{rgb}{0.7, 0.7, 0.7}
\definecolor{grey}{rgb}{0.4, 0.4, 0.4}

\def\red{\color{red}}

\begin{document}


\title{The RN/CFT Correspondence Revisited}

\author{Chiang-Mei Chen} \email{cmchen@phy.ncu.edu.tw}
\affiliation{Department of Physics and Center for Mathematics and Theoretical Physics,
National Central University, Chungli 320, Taiwan}

\author{Jia-Rui Sun} \email{jrsun@phy.ncu.edu.tw}
\affiliation{Department of Physics, National Central University, Chungli 320, Taiwan}

\author{Shou-Jyun Zou} \email{sgzou@hotmail.com}
\affiliation{Department of Physics, National Central University, Chungli 320, Taiwan}

\date{\today}


\begin{abstract}
We reconsidered the quantum gravity description of the near horizon
extremal Reissner-Nordstr{\o}m black hole in the viewpoint of the
AdS$_2$/CFT$_1$ correspondence. We found that, for pure electric
case, the right moving central charge of dual 1D CFT is $6 Q^2$
which is different from the previous result $6 Q^3$  of left moving
sector obtained by warped AdS$_3$/CFT$_2$ description. We discussed
the discrepancy in these two approaches and examined novel
properties of our result.
\end{abstract}


\maketitle
\tableofcontents

\section{Introduction}
Searching for the full quantum theory of gravity is still a tough
problem. Although the holographic principle, in particular the
explicit realization for AdS/CFT correspondence, open a new visual
angle to solve this problem \cite{'tHooft:1993gx, Susskind:1994vu,
Maldacena:1997re, Gubser:1998bc, Witten:1998qj}, there are so far
only few well-understood examples such as the AdS$_5$/CFT$_4$
\cite{Maldacena:1997re} and AdS$_3$/CFT$_2$ \cite{Brown:1986nw}
dualities. Recently, an interesting progress has been made along
this direction for extremal Kerr black holes---the Kerr/CFT
correspondence \cite{Guica:2008mu}, in which the central charge of
the dual chiral CFT (left movers) can be identified by studying the
asymptotic symmetry of the near horizon extremal Kerr black hole
(NHEK) geometry. The NHEK geometry contains an AdS$_2$ factor with
an $S^1$ bundle assembled as a warped AdS$_3$ geometry which
involves $SL(2,\mathbb{R})_R \times U(1)_L$ symmetry. This
consequence indicates that a new duality, warped AdS$_3$/CFT$_2$
correspondence, may exist. Soon after, the Kerr/CFT correspondence
has been generalized into many other spacetimes which contain a
warped AdS$_3$ structure \cite{Hotta:2008xt, Lu:2008jk,
Azeyanagi:2008kb, Chow:2008dp, Azeyanagi:2008dk, Nakayama:2008kg,
Isono:2008kx, Peng:2009ty, Chen:2009xja, Loran:2009cr,
Ghezelbash:2009gf, Lu:2009gj, Amsel:2009ev, Dias:2009ex, Compere:2009dp, Krishnan:2009tj,
Hotta:2009bm, Astefanesei:2009sh, Wen:2009qc, Azeyanagi:2009wf,
Wu:2009di, Matsuo:2009sj, Bredberg:2009pv, Cvetic:2009jn,
Hartman:2009nz}. However, the initial work \cite{Guica:2008mu}
did not address more information about the dual CFT in
addition to the central charge. Therefore, new schemes should be
developed to understand the properties of the CFT and dynamics
behind this duality. A promising suggestion is to construct the
stress tensor of the dual CFT (2D or 1D) and study the corresponding
properties from the conserved charges \cite{Amsel:2009pu,
Balasubramanian:2009bg, Castro:2009jf}.

Besides, it is also interesting to ask whether we can find the CFT
description for extremal non-rotating black holes and for near
extremal generalization. Indeed, the CFT dual of extremal
Reissner-Nordstr{\o}m (RN) black hole has been studied by using the
analogous treatment in Kerr/CFT correspondence \cite{Hartman:2008pb,
Garousi:2009zx}. In that approach, the off-diagonal part of the
metric, the $U(1)$ bundle of warped AdS$_3$, was recovered from the
gauge field potential by uplifting the RN black hole into 5D
gravity. Then the left-moving central charge, $c_L = 6 Q^3$, and
temperature, $T_L = 1/2 \pi Q$, were computed \cite{Hartman:2008pb,
Garousi:2009zx} which reproduce the black hole entropy by the Cardy
formula. However, these results have unlike properties comparing
with Kerr black hole case. For example, the CFT temperature is
dependent on the black hole parameter unlike the constant
temperature $T_L = 1/2 \pi$ for Kerr black hole. Moreover, the
central charge (left movers) in the Kerr/CFT correspondence $c_L =
12 J = 6 l^2/G_4$ is related to the angular momentum, $J$, or
equivalently to the square of the warped AdS$_3$ radius, $l$, of the
NHEK geometry \cite{Guica:2008mu}. While in the near horizon (near)
extremal RN black hole geometry, the radius of AdS$_2$ factor is
given by the charge parameter $q$. Thus a naive but reasonable
expectation is that the central charge corresponding to the RN black
hole is proportional to $q^2$ and the level-central charge relation
$h_L = c_L/24$ for AdS$_3$ could hold, then the CFT temperature is
still a constant $T_L = 1/2 \pi$. In this paper, we check such
conjecture by considering the 2D effective action of 4D extremal RN
black hole via a dimensional reduction. Analogous to the recent work
by Castro and Larsen \cite{Castro:2009jf} on the near extremal Kerr
black hole, we used the boundary counterterm method
\cite{Balasubramanian:1999re, de Haro:2000xn} in the framework of
the AdS$_2$/CFT$_1$ correspondence to derive the ``right-moving''
central charge. Our result shows that the central charge of 1D CFT
dual to the extremal RN black hole is $c_R = 6 q^2/G_4 = 6 Q^2$.
Since there is no gravitational anomaly, it is reasonable to expect
$c_L = c_R$, then using the Cardy formula, the entropy of extremal
RN black hole is reproduced, which confirms the conjecture. Note
that, if we identify $c_L$ with $c_R$, then our result indicates a
discrepancy compared with those obtained in \cite{Hartman:2008pb,
Garousi:2009zx} from the warped AdS$_3$ descriptions. Actually,
there is a scaling ambiguity in both AdS$_2$ and warped AdS$_3$
descriptions to determine the central charge dual to RN black hole.
We will discuss the details on issue in the section of Conclusion
and Discussion.

This paper is organized as follows. In section II, we briefly review
the near horizon near extremal RN black hole, and then in section
III, we study the stress tensor and current of the dual 1D CFT by
using the boundary counterterm method. In section IV, by imposing
the appropriate asymptotic boundary conditions and combing of the
diffeomorphism transformation together with the gauge
transformation, we obtain the central charge of the dual CFT and
calculate the conserved charges. In section V, we give another
derivation to support our result. Finally, we come to the conclusion
and discussion.  Moreover, the derivation of the central charge for
dyonic RN black holes is briefly summarized in the appendix A.

\section{Review of the near extremal RN black hole}
The action of the 4D Einstein-Maxwell theory
\begin{equation}\label{EMaction}
I = \frac1{16 \pi G_4} \int d^4x \sqrt{-g_4} \left( R_4 - F^2 \right),
\end{equation}
admits uniquely spherically symmetric electro-vacuum solution---the Reissner-Nordstr{\o}m (RN) black hole. The explicit expressions for the electric charged RN black hole are
\begin{eqnarray}\label{rn}
ds^2 &=& - \left( 1 - \frac{2 m}{r} + \frac{q^2}{r^2} \right) dt^2 +
\frac{dr^2}{1 - \frac{2 m}{r} + \frac{q^2}{r^2}} + r^2 d\Omega_2^2,
\nonumber\\
A &=& \frac{q}{r} \, dt, \qquad F = \frac{q}{r^2} \, dt \wedge dr,
\end{eqnarray}
in which two parameters $m, q$ representing the mass and electric charge.\footnote{The parameters $m$ and $q$ have dimension of length and the ADM mass and the electric charge (dimensionless) are given by $M = m/G_4, Q = q/\sqrt{G_4}$. }
The general RN black hole has two horizons and the corresponding outer and inner horizon radii are
\begin{equation}
r_\pm = m \pm \sqrt{m^2 - q^2}.
\end{equation}
The region enclosed by outer horizon behaves like a thermodynamical system with Hawking temperature and Bekenstein-Hawking entropy
\begin{equation}\label{TS}
T_H = \frac{\kappa}{2 \pi} = \frac{r_+ - r_-}{4 \pi r_+^2}, \qquad
S_{BH} = \frac{A_+}{4 G_4} = \frac{\pi r_+^2}{G_4},
\end{equation}
where $\kappa$ and $A_+$ are the surface gravity and area of the outer horizon, respectively.

The near horizon geometry of extremal RN black holes, i.e. $m = q$
(assuming $q > 0$), incudes a $AdS_2$ structure revealing existence
of a CFT description. Technically, one can obtain the near horizon
geometry of near extremal RN black hole by taking the following
limits of $\varepsilon \to 0$
\begin{equation}\label{trans}
r \to q + \varepsilon \rho, \qquad t \to \frac{\tau}{\varepsilon},
\qquad m = q + \frac{\varepsilon^2 b^2}{2 q},
\end{equation}
and the near horizon solution is \cite{Bertotti:1959pf, Robinson:1959ev,Maldacena:1998uz}
\begin{eqnarray}\label{nhrn}
ds^2 &=& - \frac{\rho^2 - b^2}{q^2} d\tau^2 + \frac{q^2}{\rho^2 -
b^2} d\rho^2 + q^2 d\Omega_2^2,
\nonumber\\
A &=& - \frac{\rho}{q} \, d\tau, \qquad F = \frac1{q} \, d\tau \wedge d\rho.
\end{eqnarray}
Here the radius of AdS$_2$ is given by the charge parameter $q$ and the combination
$\varepsilon b$ labels the derivation from the extremality.
Thus, the black hole entropy can be expanded as
\begin{equation}\label{entropy}
S_{BH} = \frac{\pi}{G_4} \left( q^2 + 2 q \varepsilon b + \mathcal{O}(\varepsilon^2 b^2) \right).
\end{equation}
For the extremal case, i.e. $b = 0$, the black hole entropy, $S_{BH}
= \pi \, q^2 /G_4$, is expected to match the entropy of the dual CFT
calculated from the Cardy formula
\begin{equation}\label{entropyCFT}
S_{CFT} = 2 \pi \sqrt{\frac{c_L h_L}6}.
\end{equation}

The central charge and temperature, in the extremal
limit, have been calculated in the warped AdS$_3$/CFT$_2$ picture \cite{Hartman:2008pb, Garousi:2009zx}
\begin{equation}\label{cL}
c_L = 6 Q^3, \qquad T_L = \frac1{2 \pi Q} \quad \Rightarrow
\quad h_L = \frac{\pi^2 T^2 c}6 = \frac{Q}{4}.
\end{equation}
However, the relations $c_L \propto Q^3$ and $T_L \propto 1/Q$ seem
very unnatural. Inspired from the Kerr/CFT results
\cite{Castro:2009jf}, one may expect that the central charge is
proportional to the square of AdS$_2$ radius is a general picture,
which implies  $c_L \propto q^2$ for RN black hole. Moreover, if we
further assume the relations $h_L = c_L/24$ is still hold (like the
2D CFT case), then the matching of black hole and CFT entropies
gives
\begin{equation}\label{guess}
c_L = \frac{6 q^2}{G_4} = 6 Q^2, \qquad h_L = \frac{q^2}{4
G_4} = \frac14 Q^2, \qquad T_L = \frac1{2\pi}.
\end{equation}
These expected results involve physically promising properties. In
the following sections, we will explicitly calculate the
right-moving central charge to verify our speculation
eq.(\ref{guess}) based on the expectation $c_L = c_R$ (since there
is no gravitational anomaly).

\section{The 2D effective theory}
We study the CFT dual description for the RN black hole by considering a 2D effective action dimensionally reduced from the 4D Einstein-Maxwell action (\ref{EMaction}). By assuming proper ansatz for metric and gauge potential
\begin{equation}
ds^2 = g_{\mu\nu} dx^\mu dx^\nu + q^2 \mathrm{e}^{-2\psi} d\Omega_2^2, \qquad A = A_\mu dx^\mu,
\end{equation}
we can straightforwardly compute the following geometric quantities
\begin{equation}
R_4 = R_2 + \frac2{q^2} \mathrm{e}^{2 \psi} - 2 \mathrm{e}^{2\psi} \nabla^2 \mathrm{e}^{-2 \psi}
+ 2 \mathrm{e}^{2 \psi} \nabla_\mu \mathrm{e}^{-\psi} \nabla^\mu \mathrm{e}^{-\psi},
\end{equation}
and
\begin{equation}
\sqrt{-g_4} = q^2 \mathrm{e}^{-2\psi} \sin\theta \sqrt{-g_2}.
\end{equation}
Therefore, the 2D effective action is
\begin{equation}\label{2Daction}
I = \frac{q^2}{4 G_4} \int d^2x \sqrt{-g_2} \left(
\mathrm{e}^{-2\psi} R_2 + \frac2{q^2} + 2 \nabla_\mu
\mathrm{e}^{-\psi} \nabla^\mu \mathrm{e}^{-\psi} -
\mathrm{e}^{-2\psi} F^2 \right),
\end{equation}
where $R_2$ is the 2D Ricci scalar associated with $g_{\mu\nu}$,
$\psi$ is the dilaton field and $F_{\mu\nu} = \partial_\mu
A_{\nu} - \partial_\nu A_{\mu}$ is the $U(1)$ gauge field strength.
The corresponding equations of motion for constant $\psi$, which is a consistent truncation, are (note
that $R_{\mu\nu} = \frac 1 2 R g_{\mu\nu}$ for 2D geometry)
\begin{eqnarray}
R_2 - F^2 &=& 0,
\nonumber\\
\nabla_\mu F^{\mu\nu} &=& 0,
\nonumber\\
\frac12 g_{\mu\nu} \left( \frac1{q^2} \mathrm{e}^{2\psi} - \frac12 F^2 \right) +
F_{\mu\alpha} F_\nu{}^\alpha &=& 0.
\end{eqnarray}
From the above equations we get
\begin{equation}
R_2 = F^2 = - \frac2{q^2} \mathrm{e}^{2\psi}.
\end{equation}
Therefore the radius of local AdS$_2$ solution, with constant $\psi$, is
\begin{equation}
\ell_{AdS} = q \, \mathrm{e}^{-\psi}.
\end{equation}
For the near horizon (near) extremal RN black hole,
$\mathrm{e}^{-\psi} = 1$. By a suitable choice of gauges, we assume
the 2D general solution also takes the form \cite{Castro:2009jf}
\begin{equation}\label{2dmetric}
ds^2 = g_{\mu\nu}dx^{\mu}dx^{\nu}=\mathrm{e}^{-2\psi} dr^2 +
\gamma_{tt}(t,r) dt^2, \qquad A = A_t(t,r) dt,
\end{equation}
then the equations of motion reduce to
\begin{equation}
\gamma_{tt} = - q^2 (\partial_r A_t)^2, \qquad 2 \gamma_{tt} \partial^2 \gamma_{tt} -
(\partial_r \gamma_{tt})^2 - 4 q^{-2} \gamma_{tt}^2 = 0.
\end{equation}
The general solution is characterized by a free time-dependent function
\begin{eqnarray}
\gamma_{tt} &=&
- \frac{q^2}{16} \left[ \mathrm{e}^{r/q} - f(t) \mathrm{e}^{-r/q} \right]^2 \sim - \frac{q^2}{16} \mathrm{e}^{2r/q},
\\
A_t &=&
-\frac{q}4 \mathrm{e}^{r/q} \left[ 1  - \sqrt{f(t)} \mathrm{e}^{-r/q} \right]^2 \sim -\frac{q}4 \mathrm{e}^{r/q}.
\end{eqnarray}
The special solution $f = 0$ corresponds to extremal case, and for the near extremal case $f = 4 b^2 / q^4$. Moreover, the extrinsic curvature at boundary, i.e. $r \to \infty$, is
\begin{equation}
K = \lim_{r\to\infty} \frac12 \gamma^{tt} n^\mu \partial_\mu \gamma_{tt} = \frac1{q} \mathrm{e}^\psi,
\end{equation}
where $n^r = \sqrt{g^{rr}} = \mathrm{e}^\psi$.

\subsection{Boundary counterterms}
In order to obtain the renormalized finite boundary stress tensor of
the dual 1D CFT, an effective way is to add suitable
boundary counterterms, for gravity and matter fields, into
the action \cite{Balasubramanian:1999re, de Haro:2000xn}. In the
asymptotic AdS$_2$ case considered here, the geometric counterterm is
just the Gibbons-Hawking term and the rest boundary counterterm comes
from the contribution of gauge
field, i.e.
\begin{equation}
I_\mathrm{bndy} = I_\mathrm{GH} + I_\mathrm{counter},
\end{equation}
where
\begin{eqnarray}
I_\mathrm{GH} &=& \frac{q^2}{2 G_4} \int dt \sqrt{-\gamma} \mathrm{e}^{-2\psi} K,
\\
I_\mathrm{counter} &=& \frac{q^2}{2 G_4} \int dt \sqrt{-\gamma} \left( \alpha \mathrm{e}^{-\psi} + \beta \mathrm{e}^{-\psi} A_a A^a \right),
\end{eqnarray}
and $\alpha, \beta$ are constants to be determined. The variation of
full action takes the form
\begin{equation}
\delta I = (\mathrm{bulk \ terms}) + \int dt \sqrt{-\gamma} \left(
\pi_{ab} \delta \gamma^{ab} + \pi_\psi \delta\psi + \pi^a \delta A_a \right),
\end{equation}
where
\begin{eqnarray}
\pi_{tt} &=&  - \frac{q^2}{4 G_4} \left( \alpha \mathrm{e}^{-\psi}
\gamma_{tt} + \beta \mathrm{e}^{-\psi} A_a A^a \gamma_{tt} -2 \beta
\mathrm{e}^{-\psi} A_t A_t \right) \sim - \frac{q^2}{4 G_4}
\mathrm{e}^{-\psi} \gamma_{tt} ( \alpha + \beta ),
\nonumber\\
\pi_\psi &=& - \frac{q^2}{2 G_4} \left( 2 \mathrm{e}^{-2\psi} K + \alpha \mathrm{e}^{-\psi} + \beta \mathrm{e}^{-\psi} A_a A^a \right) \sim - \frac{q^2}{2 G_4} \mathrm{e}^{-\psi} \left( \frac2{q} + \alpha - \beta \right),
\nonumber\\
\pi^t &=&  \frac{q^2}{4 G_4} \left( - 4 \mathrm{e}^{-2\psi} n_\mu
F^{\mu t} + 4 \beta \mathrm{e}^{-\psi} A^t \right) \sim \frac{q^2}{4
G_4} \mathrm{e}^{-\psi} \gamma^{tt} \mathrm{e}^{r/q} ( 1 - \beta q
).
\end{eqnarray}
A well-defined variational principle requires all the coefficients
$\pi_{ab}$, $\pi_\psi$ and $\pi^a$ be finite. Similar to the Kerr
black hole case considered in \cite{Castro:2009jf}, the values of
$\alpha, \beta$ can be determined by ensuring vanishing leading
terms of these three boundary momenta, and we obtain
\begin{equation}
\alpha = - \frac1{q}, \qquad \beta = \frac1{q}.
\end{equation}

\subsection{Boundary currents}
The boundary currents are defined by
\begin{eqnarray}\label{current}
T_{ab} &=& - \frac{2}{\sqrt{-\gamma}} \frac{\delta I}{\delta
\gamma^{ab}} = - 2 \pi_{ab},
\nonumber\\
J^a &=& - \frac1{\sqrt{-\gamma}} \frac{\delta I}{\delta A_a} = - \pi^a,
\end{eqnarray}
and their corresponding components are
\begin{eqnarray}\label{current1}
T_{tt} &=&  - \frac{q}{2 G_4} \mathrm{e}^{-\psi} ( \gamma_{tt} + A_t
A_t ) = \frac{q^3}{8 G_4} \mathrm{e}^{-\psi} \mathrm{e}^{r/q}
\sqrt{f} \left( 1 - \sqrt{f} \mathrm{e}^{-r/q} \right)^2,
\nonumber\\
J_t &=&  - \frac{q}{G_4} \mathrm{e}^{-\psi} \left( A_t - q
\mathrm{e}^{-\psi} n^\mu F_{\mu t} \right) = - \frac{q^2}{2 G_4}
\mathrm{e}^{-\psi} \sqrt{f} \left( 1 - \sqrt{f} \mathrm{e}^{-r/q}
\right).
\end{eqnarray}
The exact solution corresponding to near horizon of near extremal RN black hole is
\begin{eqnarray}\label{2drn}
\gamma_{tt} &=& - \frac{q^2}{16} \left( \mathrm{e}^{r/q} - \frac{4
b^2}{q^4} \mathrm{e}^{-r/q} \right)^2,
\nonumber\\
A_t &=& - \frac{q}4 \mathrm{e}^{r/q} \left( 1  - \frac{2 b}{q^2} \mathrm{e}^{-r/q} \right)^2,
\end{eqnarray}
and consequently the associated currents are
\begin{eqnarray}
T_{tt} &=& \frac{q b}{4 G_4} \mathrm{e}^{-\psi} \left(
\mathrm{e}^{r/q} - \frac{4 b}{q^2} + \frac{4 b^2}{q^4} \mathrm{e}^{-
r/q} \right),
\nonumber\\
J_t &=&  - \frac{q^2}{G_4} \mathrm{e}^{-\psi} \left( \frac{b}{q^2} -
\frac{2b^2}{q^4} \mathrm{e}^{-r/q} \right).
\end{eqnarray}
Like the near extremal Kerr black hole case, for $r \gg q$, the $T_{tt}$ is still divergent.
\omits{A possible explanation is that we missed some counterterm
such as the interacting term of the form $KA^a_A_a$, i.e., the
minimal coupling term $A^aA_a$ is not enough to cancel the
divergence.}

\section{Asymptotic symmetries and conserved charges}
As has been point out in \cite{Hartman:2008dq, Castro:2009jf}, for a
chosen asymptotic symmetry of spacetime, the gauge fields need not
only be diffeomorphism invariant but also be gauge invariant. We can
see that the combination of these two transformations indeed give
the central charge we expected. The asymptotic boundary conditions
are
\begin{equation}
\delta_\epsilon g_{rr} = 0, \qquad \delta_\epsilon g_{tr} = 0, \qquad \delta_\epsilon g_{tt} = 0 \cdot \mathrm{e}^{2r/q} + \cdots,
\end{equation}
which lead to the allowed transformations
\begin{equation}
\epsilon^r = - q \partial_t \xi(t), \qquad \epsilon^t = \xi(t) + 8 \mathrm{e}^{-2\psi} \left( \mathrm{e}^{2r/q} - f(t) \right)^{-1} \partial_t^2 \xi(t).
\end{equation}
The boundary metric is transformed as
\begin{equation}
\delta_\epsilon \gamma_{tt} = q^2 \left( 1 - f \,
\mathrm{e}^{-2r/q} \right) \left[ \frac14 f \partial_t \xi + \frac18 \xi \partial_t f - \mathrm{e}^{-2\psi} \partial_t^3 \xi \right].
\end{equation}
However, the result
\begin{equation}
\delta_\epsilon A_r = 4 \partial_t^2 \xi \, \mathrm{e}^{-2\psi} \mathrm{e}^{-r/q} \left( 1 + \sqrt{f} \, \mathrm{e}^{-r/q} \right)^{-2},
\end{equation}
violates the gauge condition $A_r = 0$. So we need a compensated gauge transformation
\begin{equation}
\Lambda
= 4 q \mathrm{e}^{-2\psi} \mathrm{e}^{-r/q} \left( 1 + \sqrt{f} \, \mathrm{e}^{-r/q} \right)^{-1} \partial_t^2 \xi,
\end{equation}
such that
\begin{equation}
\delta_{\epsilon + \Lambda} A_r = \delta_\epsilon A_r + \partial_r \Lambda = 0.
\end{equation}
Thus variation of the time component of gauge field is
\begin{eqnarray}
\delta_{\epsilon + \Lambda} A_t &=& \frac{q}2 \left( 1 - \sqrt{f} \, \mathrm{e}^{-r/q} \right) \partial_t \left( \sqrt{f} \, \xi \right) + 2 q \mathrm{e}^{-2\psi} \mathrm{e}^{-r/q} \partial_t^3 \xi
\nonumber\\
&=& \frac{q}2 \partial_t \left( \sqrt{f} \, \xi \right) + \frac{q}2 \mathrm{e}^{-r/q} \left( - f \partial_t \xi - \frac12 \xi \partial_t f + 4 \mathrm{e}^{-2\psi} \partial_t^3 \xi \right),
\end{eqnarray}
and variation of the stress tensor is
\begin{eqnarray}
\delta_{\epsilon + \Lambda} T_{tt} &=&  - \frac{q}{2 G_4}
\mathrm{e}^{-\psi} ( \delta_\epsilon \gamma_{tt} + 2 A_t
\delta_{\epsilon + \Lambda} A_t )
\nonumber\\
&=& 2 T_{tt} \partial_t \xi + \xi \partial_t T_{tt}
\nonumber\\
&& + \frac{q^3}{G_4} \left[ \mathrm{e}^{-3\psi} \partial_t^3 \xi -
\frac18 \mathrm{e}^{-\psi} \sqrt{f} \, \mathrm{e}^{r/q} \left( 1 +
\sqrt{f} \, \mathrm{e}^{-r/q} \right) \partial_t \xi \right] \left(
1 - \sqrt{f} \, \mathrm{e}^{-r/q} \right).
\end{eqnarray}
The right-moving central charge can be read out from the
following relation \cite{Castro:2008ms}
\begin{equation}
\delta_{\epsilon + \Lambda} T_{tt} = 2 T_{tt} \partial_t \xi + \xi \partial_t T_{tt} - \frac{c}{12} L \partial_t^3 \xi,
\end{equation}
here the normalization factor $L$ of dimension of length is needed
to ensure the central charge to be dimensionless. It turns out that
a physically suitable value of the normalization is proportional to
the AdS radius $L$ by a factor $-2$, namely $L = - 2 \, \ell_{AdS} =
- 2 q \mathrm{e}^{-\psi}$, due to the same factor appearing in the
cosmological constant term in the effective action (\ref{2Daction}).
\footnote{The effective AdS radius indeed is exactly the notion $L$
defined in the action (\ref{HSaction}).} Actually, in the next
section, we will see that this factor $-2$ endorses the ground value
for the level. Finally we can easily read out the right moving
central charge for the extremal RN black hole ($f = 0, \psi = 0$)
\begin{equation}\label{cL1}
c_R = \frac{3 L^2}{2 G_4} = \frac{6 q^2}{G_4} = 6 Q^2.
\end{equation}
Since there is no gravitational anomaly, similar to the near
extremal Kerr black hole case \cite{Castro:2009jf}, it is reasonable
to expect that $c_L = c_R$, This is just we have expected in
eq.(\ref{guess}).

\omits{\red The Noether charge generated by the infinitesimal
diffeomorphism transformation has only the time-time component,
which is associated with the excited energy of the right movers
{\red (we should explain why minus sign)}
\begin{equation}
Q_\epsilon = T_{tt} - J_t A_t = - \frac{q^3}{8 G_4} f \mathrm{e}^{-\psi} \left( 1 - \sqrt{f} \mathrm{e}^{-r/q} \right)^2 \approx - \frac{q^3}{8 G_4} f \mathrm{e}^{-\psi}.
\end{equation}
For the near extremal RN, $Q_\epsilon = - b^2/(2 G_4 q)$, which, up
to a convention sign, measures the excitation energy in the ADM
energy
\begin{equation}
M = \frac{m}{G_4} = \frac{q}{G_4} + \varepsilon^2 \frac{b^2}{2 G_4 q}.
\end{equation}
}

For the near horizon extremal RN black hole, the Noether
charge generated by the gauge transformation is
\begin{equation}
\delta_\Lambda I =  - \int dt \sqrt{-\gamma} \mathcal{J}^a
\partial_a \Lambda,
\end{equation}
therefore
\begin{equation}
\mathcal{J}^t = - \frac{\delta I_\mathrm{counter}}{\sqrt{-\gamma}
\delta A_t} = - \frac{q}{G_4} \mathrm{e}^{-\psi} A^t,
\end{equation}
and the associated charge is
\begin{equation}
Q_\Lambda = \sqrt{-\gamma} \mathcal{J}^t = \frac{q \mathrm{e}^{-\psi}}{G_4} \frac{1 - \sqrt{f} \mathrm{e}^{-r/q}}{1 + \sqrt{f} \mathrm{e}^{-r/q}} \approx \frac{q}{G_4} \mathrm{e}^{-\psi},
\end{equation}
which identical to the ADM energy for the extremal RN black hole.

The level $k$ is defined as
\begin{equation}
\delta_\Lambda \mathcal{J}_t = \frac{k}2 L \, \partial_t \Lambda,
\end{equation}
and
\begin{equation}
\delta_\Lambda \mathcal{J}_t = - \frac1{2 G_4} (2 q \mathrm{e}^{-\psi}) \partial_t \Lambda,
\end{equation}
therefore
\begin{equation}
k = \frac1{G_4}.
\end{equation}
Note that the presence of the gravitational constant $G_4$ is due
the the overall factor $1/16\pi G_4$ in the matter part of the
action (\ref{EMaction}). In the usual convention, this factor is
absorbed in the Maxwell field strength and then the value is $k =
1$. Moreover, the  normalization factor $L$ in the definition of
level ensures the ground value.

\section{Another derivation}
In this section we will check the validity of the central charge eq.(\ref{cL1})
by comparing with the results in \cite{ Hartman:2008dq, Castro:2008ms}.
Recall that our effective action (\ref{2Daction}) is consistently
truncated, for $\psi = 0$, to
\begin{equation}
I = \frac1{4 G_4} \int d^2x \sqrt{-g_2} \left[ q^2 \left( R_2 +
\frac2{q^2} \right) -  q^2 F^2 \right],
\end{equation}
which is equivalent to the action considered in \cite{Castro:2008ms}
\begin{equation} \label{HSaction}
I = \kappa \int d^2x \sqrt{-g_2} \left[ \eta \left(
R_2 + \frac8{L^2} \right) -  \frac{L^2}4 F^2 \right],
\end{equation}
by simply identifying
\begin{equation}
\kappa = \frac1{4 G_4}, \qquad L = - 2 q, \qquad \eta = q^2 = \frac{L^2}4.
\end{equation}
The central charge for (\ref{HSaction}) was derived in \cite{Hartman:2008dq, Castro:2008ms} as
\begin{equation}
c = \frac32 k L^4 E^2,
\end{equation}
where the electric field is $E^2 = 4 \eta / L^4 = 1 / L^2$. It's
easy to see that our results of the central charge, $c = 6 q^2/G_4 =
3 L^2 / 2 G_4$ and level $k = 1/G_4$ are consistent with this
relation. Indeed our results for level and right moving central
charge can be directly computed via the formulae in
\cite{Castro:2008ms} (with an opposite sign due to the negative
value of $L$)
\begin{equation}\label{kandc}
k = 4 \kappa, \qquad c = 24 \kappa \eta.
\end{equation}

Our results indicate that the central charge derived from
the viewpoint of AdS$_2$/CFT$_1$ correspondence is the same as the
one derived from the 2D CFT on a strip. This equality seems hint
that CFT$_1$ can be regarded as an chiral CFT$_2$ \cite{Sen:2008yk, Alishahiha:2008tv, Gupta:2008ki}.

\section{Conclusion and discussion}
In this paper, we reconsidered the quantum gravity description of
near horizon extremal RN black hole by using the boundary
counterterm approach in the context of AdS$_2$/CFT$_1$
correspondence. We found that the right moving central charge and
the level of the dual 1D CFT are $c_R = 6 q^2/G_4 = 6 Q^2$ and $k =
1/G_4$, respectively. Since there is no gravitational
anomaly, similar to the near extremal Kerr black hole case
\cite{Castro:2009jf}, it is naturally to expect that the left and
right moving central charges are identical. Therefore it seems to
have a discrepancy comparing with the previous result of left moving
central charge $c_L = 6  Q^3$ obtained from the warped
AdS$_3$/CFT$_2$ prescription in \cite{Hartman:2008pb,
Garousi:2009zx}.Regarding to the result i.e. $c_L =
6 Q^3$ in \cite{Hartman:2008pb, Garousi:2009zx}, we noticed that
authors uplifted the RN black hole into higher dimensions by simply
assuming the radius of extra dimensional cycle to be one and its
period to be $2\pi$. In this approach, however, the size of the
extra dimension could affect the result of the central charge and
there are no physically sensible preference to determine this size.
Although, there is also a normalization scale of AdS radius
(negative) to be specified in our approach, but we have more clear
physical prescriptions to resolve this scale. Therefore, the result
of central charge is supported by several novel properties.
Moreover, as a crosscheck, our central charge and level agree with
the results obtained from a 2D Maxwell-dilaton quantum gravity which
is equivalent to a CFT$_2$ on a strip \cite{Hartman:2008dq}. The
agreement of the results from different viewpoints also indicates
that CFT$_1$ may can be treated as a chiral part of CFT$_2$.

\begin{appendix}
\section{Central charge for Dyonic black holes}
We can easily generalized our analysis to the RN black holes including magnetic charge. The corresponding 2D effective action is
\begin{equation}\label{2Daction1}
I = \frac{q^2+p^2}{4 G_4} \int d^2x \sqrt{-g_2} \left( \mathrm{e}^{-2\psi} R_2 + \frac2{q^2 + p^2} - \mathrm{e}^{2\psi} \frac{2p^2}{(q^2 + p^2)^2} + 2 \nabla_\mu \mathrm{e}^{-\psi} \nabla^\mu \mathrm{e}^{-\psi} - \mathrm{e}^{-2\psi} F^2 \right).
\end{equation}
The magnetic charge parameter, $p$, adjusts the AdS$_2$ radius and
the cosmological constant term. One can derive the right moving central charge by repeating the same calculation or simply by
the relations eq.(\ref{kandc}) with the following values read out
from identifying two actions (\ref{2Daction1}) (taking $\psi = 0$)
and (\ref{HSaction})
\begin{equation}
\kappa = \frac1{4 G_4} \frac{q^2}{q^2 + p^2}, \qquad \eta =
\frac{L^4}4 = \frac{(q^2 + p^2)^2}{q^2}.
\end{equation}
Finally, the central charge is
\begin{equation}
c_R = \frac{6}{G_4} (q^2 + p^2) = 6 (Q^2 + P^2).
\end{equation}
One novel property is that the central charge also exhibits the
electric-magnetic duality which can be apparently seen from the
gravity side. However, technically, it is subtle to compute the
central charge directly for pure magnetic RN black hole. It's easy
to see that when imposing $\psi = 0$, the effective action
(\ref{2Daction1}) becomes pure 2D gravity, i.e. the cosmological
constant vanishes and the gauge field disappears. In such
circumstance, the effective AdS radius diverges and the level
vanishes.

\end{appendix}

\section*{Acknowledgement}
We would like to thank H. Lu and J.-F. Wu for very valuable
discussions. We also thank the referee for very valuable
comments on the earlier version of our paper. This work was
supported by the National Science Council of the R.O.C. under the
grant NSC 96-2112-M-008-006-MY3 and in part by the National Center
of Theoretical Sciences (NCTS).


\end{document}